\documentclass[10pt,conference]{IEEEtran} 
\IEEEoverridecommandlockouts
\usepackage{cite}
\usepackage{amsmath,amssymb,amsfonts}
\usepackage{algorithmic}
\usepackage{graphicx}
\usepackage{textcomp}
\usepackage{xcolor}
\usepackage{comment}
\usepackage{multirow}
\usepackage{url}
\def\BibTeX{{\rm B\kern-.05em{\sc i\kern-.025em b}\kern-.08em
    T\kern-.1667em\lower.7ex\hbox{E}\kern-.125emX}}
    
\newcommand{\sectopic}[1]{\vspace{0.2em}\par\noindent{\textit{\bfseries #1}}}
    
\begin{document}

\title{Towards Human-Centred Crowd Computing: Software for Better Use of Computational Resources}

\author{
    \IEEEauthorblockN{Niroshinie Fernando\IEEEauthorrefmark{1},Chetan Arora\IEEEauthorrefmark{1}\IEEEauthorrefmark{2}, Seng W. Loke\IEEEauthorrefmark{1}, Lubna Alam\IEEEauthorrefmark{1}, Stephen La Macchia\IEEEauthorrefmark{1}, Helen Graesser\IEEEauthorrefmark{1}
    } 
    \IEEEauthorblockA{\IEEEauthorrefmark{1}Deakin University, Geelong, Australia}
    \IEEEauthorblockA{\IEEEauthorrefmark{2}Monash University, Victoria, Australia}
    Email:\{niroshinie.fernando, chetan.arora, seng.loke, lubna.alam, stephen.lamacchia, h.graesser\}@deakin.edu.au
    
}


\maketitle

\begin{abstract}
Internet-connected smart devices are increasing at an exponential rate. These powerful devices have created a yet-untapped pool of idle resources that can be utilised, among others, for processing data in resource-depleted environments. The idea of bringing together a pool of smart devices for ``crowd computing'' (CC) has been studied in the recent past from an infrastructural feasibility perspective. However, for the CC paradigm to be successful, numerous socio-technical and software engineering (SE), specifically the requirements engineering (RE)-related factors are at play and have not been investigated in the literature. In this paper, we motivate the SE-related aspects of CC and the ideas for implementing mobile apps required for CC scenarios. We present the results of a preliminary study on understanding the human aspects, incentives that motivate users, and CC app requirements, and present our future development plan in this relatively new field of research for SE applications.
\end{abstract}


\section{Introduction}
The number of Internet-connected devices, such as smartphones, drones and robots, is predicted to be more than threefold the global population by 2030~\cite{VailsheryStatista2022}. Devices ``in our pockets'' and IoT devices in our living spaces are becoming ubiquitous as well as increasingly more powerful and connected, leading to a growing pool of idle resources that are not fully utilised~\cite{Arslan2014, Anderson14supercomputer}. 
These idle resources provide an exciting opportunity to be used as a distributed resource to complement infrastructure-based cloud and edge servers~\cite{Mehrabi2019,Ferrer2019decentralisedCloud, Fernando2013MccSurvey}.
Essentially, computing tasks can be ‘crowdsourced’ to the pool of human-machine resources already available within the system environment, giving rise to a new computing paradigm called ‘Crowd Computing’ (CC)~\cite{Murray2010CC, Pramanik2019CC}. CC is a socio-technical paradigm involving the technical infrastructure (e.g., cloud, edge servers and mobile devices), the human resources (e.g., the users), and the software applications (e.g., the applications for coordinating CC tasks). CC resources share and collaborate to support high-velocity and real-time jobs that are difficult to be managed by a single user/device, e.g., collaborative tasks in application areas such as tactile internet, autonomous vehicles, digital health and environmental monitoring.


CC presents numerous socio-technical and requirements engineering (RE) challenges to better use the vast (often unused or idle) resources in our smartphones and tablets in an era where computing resources are needed and sustainability is not merely an option. Technological facilitation and human motivation are needed to help us, as a society, to make the most of the computational resources already at hand, which calls for not just technical capability but also human-centred considerations in the design of CC software and systems.
Aspects of CC's technical feasibility (from an infrastructure perspective) have been demonstrated in the literature in projects such as MMPI~\cite{Doolan2008MMPI}, Hyrax~\cite{Marinelli2009Hyrax}, MClouds~\cite{Miluzzo2012mClouds}, Aura~\cite{Hasan2018Aura}, and Honeybee~\cite{Fernando2013Honeybee, Fernando2016Honeybee, Nagesh22Honeybee}. However, there are ‘human’ and software engineering (SE) aspects that have not been investigated in the CC paradigm, especially for building the CC-mobile apps. 
Based on the anecdotal evidence, the CC mobile apps' design significantly impacts the adoption of the CC paradigm among users and consequently on the overall success. Numerous ethical dilemmas, human-centred requirements and other non-functional requirements categories applicable have not been explored in the CC paradigm research.
For example, what incentivises the users to agree to share their mobile resources with other users in a CC paradigm, or which design features are most relevant to improve the probability of success in CC tasks? The existing research in requirements engineering (RE) and software solutions adoption can be key for studying CC human factors~\cite{kammuller2017security,silva2019requirements,hidellaarachchi2021effects}. The key research questions (RQs) that we aim to address in this research area are:\\
\textbf{RQ1.} What are the human-centric aspects and motivations for the adoption of the CC paradigm among CC users? \\
\textbf{RQ2.} What are the key user and system functional and non-functional requirements for designing CC mobile apps? \\
\textbf{RQ3.} Which design features and methods are appropriate for implementing the human-centric aspects (from RQ1) and requirements (from RQ2)?

In this paper, we propose our research ideas on designing CC mobile apps and RE for CC and (partially) address RQ1 and RQ2.
Below, we provide a motivating example to discuss the CC paradigm's RE and app design issues.





\begin{figure}[!t]
\centering
\includegraphics[width=0.4\textwidth]{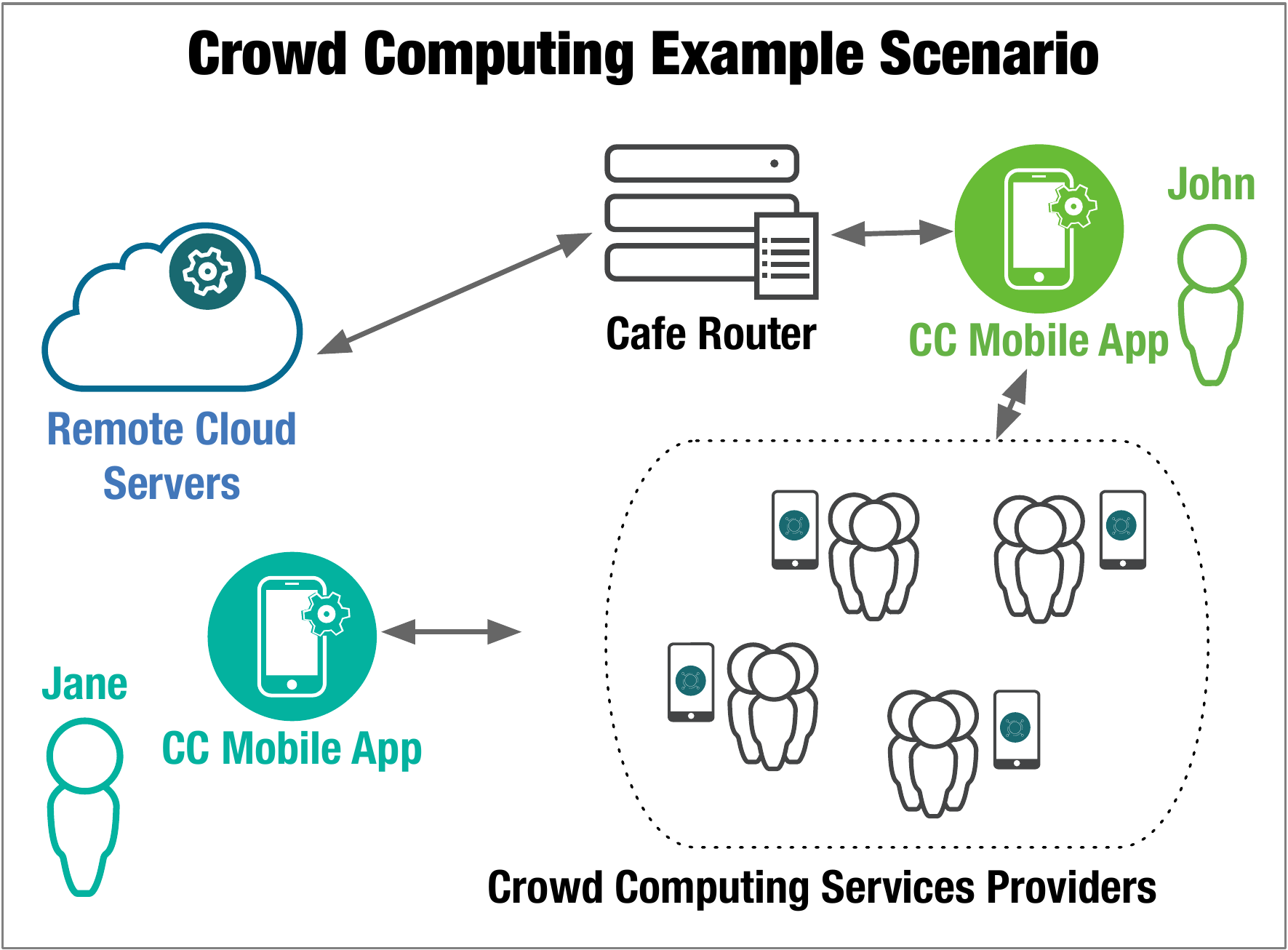}
\caption{Crowd Computing Example Scenario.}
\label{fig:cc_scenario}
\vspace*{-.5em}
\end{figure}

\sectopic{Motivating Example.} Let us consider the following scenario illustrated in Fig.~\ref{fig:cc_scenario}. Jane has a medical condition requiring her to use a wearable device to monitor her health constantly. The wearable sends the sensor data to her smartphone for analysis, but the smartphone is not powerful enough to run the artificial intelligence (AI) algorithms required to process this high-velocity stream of data continuously. Hence this health application requires the support of external resources to provide accurate and timely results (\emph{the task}). Jane considers offloading the data processing tasks directly to a remote cloud server; however, this has issues with latency and bandwidth~\cite{Shi2016EdgeComputing}, as well as high data access fees. 
In a CC paradigm, Jane would have access to computing resources of the smartphones belonging to her fellow café visitors to process her data, in effect forming a \emph{`computation crowd'}. As shown in Fig.~\ref{fig:cc_scenario}, the \emph{`computation crowd'} is now supporting Jane and another user (John).
This crowd of local device resources provides the urgently required computing resources to Jane and John, provides a faster response time and can also reduce energy use~\cite{Shi2016EdgeComputing}. Rather than utilising edge or remote resources, which are not always available in public or remote spaces, utilising the existing device resources from the crowd provides a more sustainable solution. 

Crowd computing, as illustrated in the scenario, gives rise to a number of challenges for CC app developers, from an RE perspective, as they need to take the human aspects, their motivations and the technical issues into consideration. For instance, \textit{what incentives might have convinced Jane’s fellow café visitors to share their smartphones’ computing capacity?} (information required for RQ1). There is a range of options, ranging from purely financial incentives such as micropayments directly from Jane or the café, discounts on coffee from the café, or altruistic incentives where they `donated' their devices' computing cycles to help Jane. Furthermore, \textit{what characteristics of the CC resource consumers (e.g. Jane and John) and CC resource providers (e.g., café visitors) are feasible for making this work?} (information required for RQ1). Or, \textit{how does the CC mobile app choose the best mix of computing resources to match Jane's preferences, task success guarantee, and satisfy the QoS requirements of the application?} (information required for RQ2). It is paramount that the CC applications be designed with incentive options that satisfy the requirements and priorities of both Jane (resource consumer) and the device sharers in the cafe (resource providers) to motivate uptake and participation. Lastly, \textit{how shall we design the app with hundreds of CC resource consumers and providers?} (information required for RQ3). Since the app requires the participation of hundreds of users as consumers or providers, with an equivalent high number of tasks and resource requests for different reasons, what aspects of app design be considered to ensure the successful participation of users and meeting task goals? Therefore, for the app to work, it needs to take the characteristics, motivations, and functional and non-functional requirements of users in the CC paradigm.


We present a preliminary study done via a survey to understand the requirements and incentive mechanisms of different CC users, partially addressing RQ1 and RQ2. Our results contribute to the current body of knowledge on designing apps for CC and the factors that can lead to the successful adoption of the CC paradigm. This paper further opens up avenues for future research on CC apps development that can facilitate the adoption of CC in different application domains and scenarios.

\begin{table}[!t]
\centering
\caption{CC applications in the survey with scenarios mapped to incentives}
\label{tab:survey_scenarios}
\includegraphics[width=0.45\textwidth]{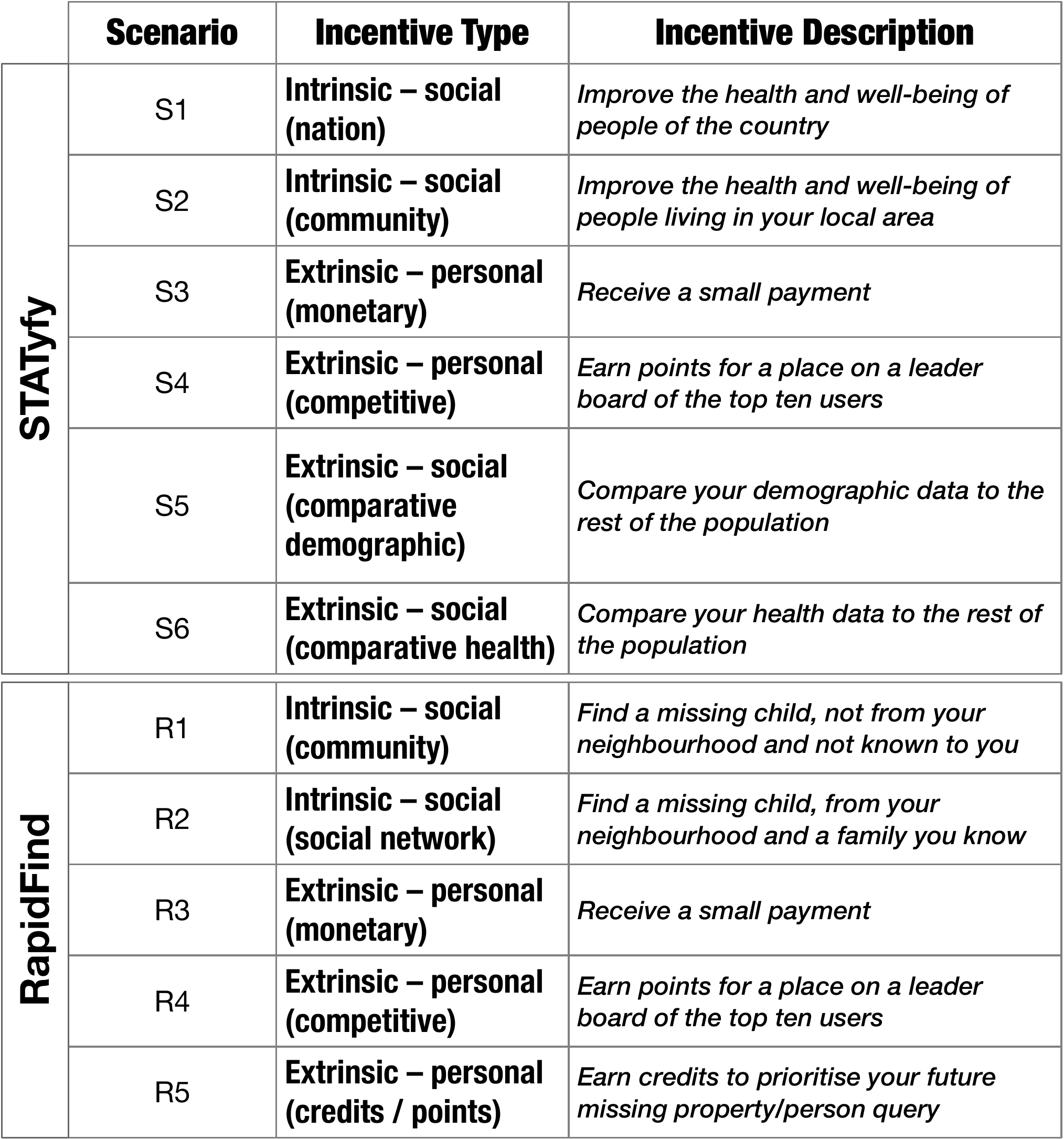}
\vspace*{-1em}
\end{table}

\section{Research Design}
We conducted an online survey consisting of different scenarios for two mock-up CC applications mapped to different incentives. The survey participants were given the description of the both apps:\\
\textbf{App1 - STATyfy}: \textit{App by the National Bureau of Statistics (NBS) that collects and analyses statistics on economic, population, environmental, and social issues. For a resource provider, STATyfy downloads data from the NBS database and performs data processing on the phone to calculate statistics on health conditions such as obesity, mental health, smoking, alcohol consumption, exercise, and food habits.} \\
\textbf{App2 - RapidFind}: \textit{App by the local police to rapidly find information about lost property and missing persons. One day, the RapidFind app requests that you participate in a CC task to find a child who's gone missing in the shopping mall on that day. If you agree, it will download a list of photos taken at the shopping mall on that day and run an image processing program, searching for the missing child. }

The survey focused on two main concerns: (1)~how likely one was to participate in a given CC scenario (uptake),  and (2)~how likely one was to frequently participate in the CC scenario (frequency) by sharing the computation power of their smartphone. 
A scenario-based methodology was used in our survey since CC is a novel computing paradigm, with only a few similar applications in the app stores and most of them being volunteer-based scientific applications, such as HTC Power To Give~\cite{htcpowertogive2016}. Therefore, we did not want to assume any prior knowledge of using CC applications among our participants. Table~\ref{tab:survey_scenarios} shows the 11 scenarios covered in our survey, S1--S6 for STATyfy and R1--R5 for RapidFind. The scenarios and incentives were derived from existing crowdsourcing literature~\cite{jaimes2015survey,wang2017mobile,phuttharak2018review}, and using the Motive-Incentive-Activation-Behavior Model (MIAB)~\cite{Leimeister2009MIAB} as a theoretical lens.

Two single-item intention measures were used as dependent variables to compare the incentive scenarios for each app. They appeared directly underneath each scenario and were answered on a seven-point scale. The uptake willingness prompt asked the participants \textit{Based on this scenario, how likely would you be to install and run [STATyfy/ RapidFind] on your smartphone?}. The seven-point scale for this prompt was: (1~=~Very unlikely, 2--4 = Neither likely nor unlikely, and 5--7 = Very likely). Following this, the intended usage frequency item prompted, \textit{Based on this scenario, how frequently would you run [STATyfy/ Rapid Find] on your smartphone?}. The scale for the frequency prompt was: (1 = Never, 2 = Once or twice at all or per year, 3 = Several times a year, 4 = 1-3 times a month, 5 = 1-3 times a week, 6 = Most days, 7 = Every day). In addition to the scenarios and two prompts on uptake and frequency, the survey also contained questions about the participants’ demographics, experiences with current volunteer-based scientific applications, proficiency with digital technology, and concerns and preferences about CC. Each survey participant was presented with all scenarios, one by one, in a randomised order for each application. The participants also had an option to add comments and feedback on CC apps, features, and inputs on the scenario.
See [link]\footnote{\url{https://nfernando.org/wp-content/uploads/2023/01/Final-Survey.pdf}} for the full wording of each scenario and the survey questions.




\begin{table}[!t]
\centering
\caption{Results of Scenario Survey}
\label{tab:surveyResults}
\includegraphics[width=0.35\textwidth]{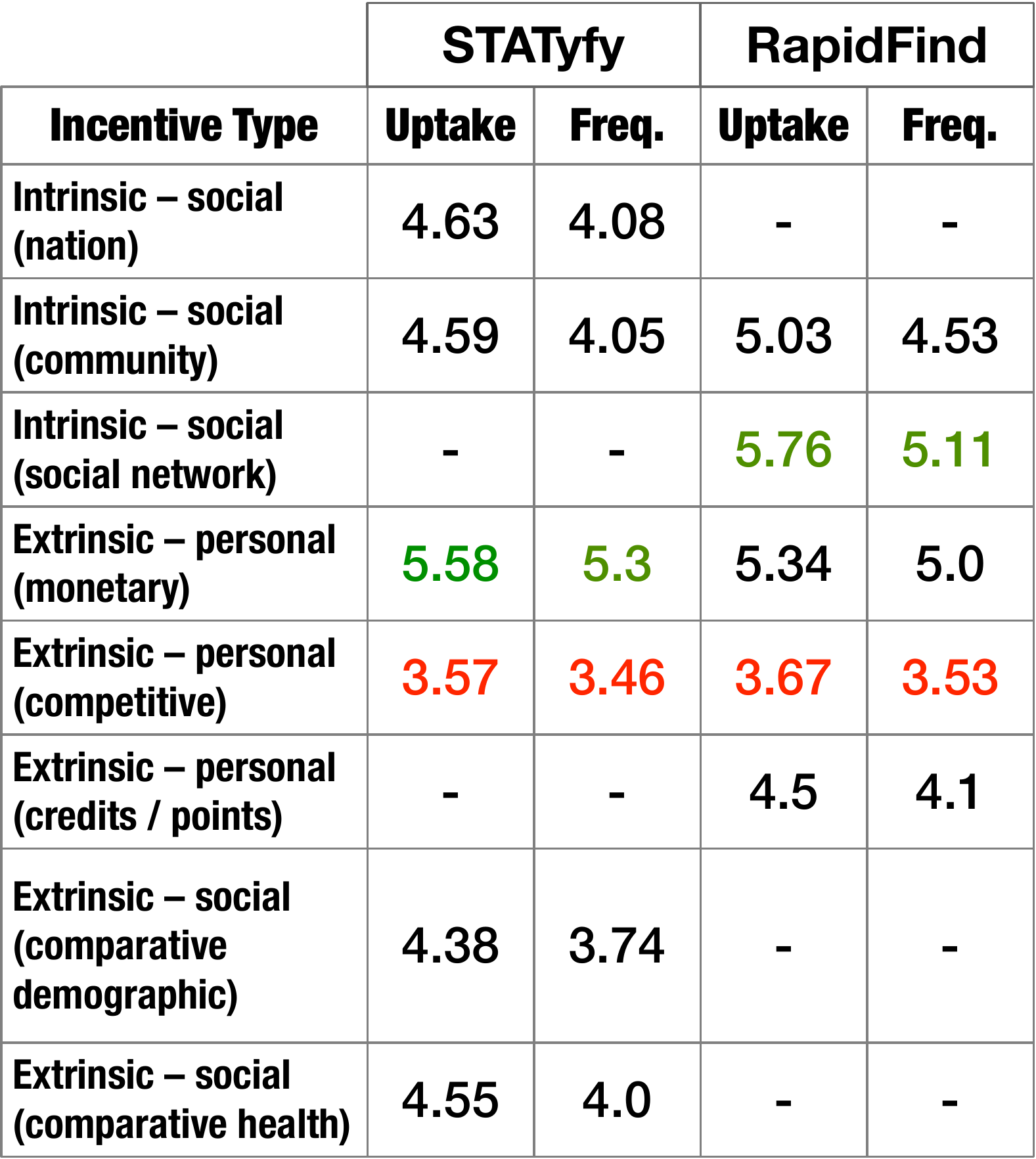}
\vspace*{-1em}
\end{table}

\section{Results and Discussion}
Table~\ref{tab:surveyResults} summarises our survey results. Overall, 219 participants responded to our survey. The table shows both apps' average uptake and frequency responses (on the seven-point scale).
In STATyfy app, scenario S3 (see Table~\ref{tab:survey_scenarios}) with extrinsic-personal (monetary payment) incentive has the highest mean response from the participants for likely uptake (significant p$<$0.05) and the frequency of uptake (significant p$<$0.05). In STATyfy, an app with no inherently attached social motivation or incentive, the monetary benefits were rated the highest in terms of incentives for people to install the app and also continue using it.
In the RapidFind app scenarios, scenario R2 (see Table~\ref{tab:survey_scenarios}) with intrinsic-social (social network) incentive had the highest mean score for likely uptake (significant p$<$0.05) and frequency of uptake (significant p$<$0.05). The results highlight that although monetary and non-monetary incentives motivate people to participate in CC when the CC application is tied to intrinsic-social incentives (e.g., to help a vulnerable person or a child), uptake can be higher than monetary incentives.


We collated the results discussed above and the qualitative feedback from our survey participants to understand the human-centric aspects and motivations for adopting the CC paradigm among CC users (RQ1) and any functional or non-functional requirements for CC apps (RQ2). We did not get any results from RQ3 in this survey. We note that even though we got some results for RQ1 and RQ2, we do not expect these to be exhaustive results for our overarching RQs for our research project. Still, they are very beneficial for directing future studies in our research project.

\begin{figure}[!t]
\centering
\includegraphics[width=0.4\textwidth]{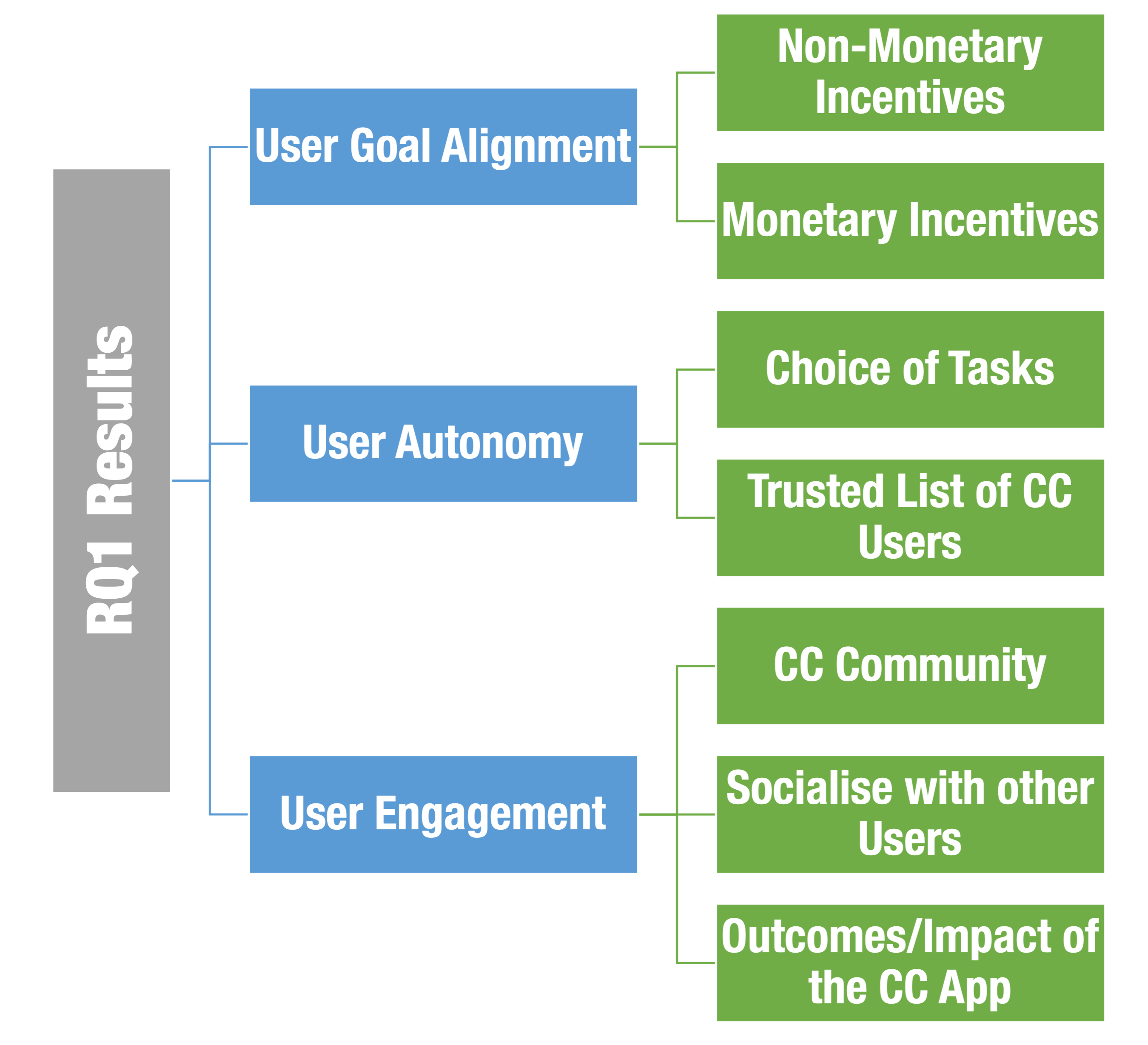}
\caption{Human-centric aspects and motivation for adopting CC paradigm.}
\label{fig:rq1}
\vspace*{-1em}
\end{figure}

Fig.~\ref{fig:rq1} shows the results of our data analysis for RQ1. We identified three key aspects in our RQ1 analysis, user goal alignment, user autonomy, and user engagement.
\textbf{User goal alignment} refers to what motivates users to participate in the CC paradigm. Findings from our initial study indicate that if user goals are strongly aligned with the CC app goal, non-monetary incentives such as altruism can work well, and monetary incentives are not required. For example, in our survey, RapidFind's goal was intrinsically aligned with helping a child. Here, the incentive with the highest uptake and continuation was a scenario with altruistic non-monetary incentives (R2). However, if the app goal is not particularly aligned with, or only weakly aligned with, user motivations, then monetary benefits such as micro-payments and credits were shown to be effective. For example, the STATyfy app's goal by itself was not intrinsically aligned with an incentive. Here, the results indicate that monetary incentives were the most effective. \textbf{User autonomy} refers to how much choice CC users have in terms of who they share resources and tasks with and what tasks they take on. Participant comments from our study revealed that having the power to only share their computing resources with a trusted list of CC users was important in some situations. \textbf{User engagement} refers to how users interact with the CC application and other users. Based on the comments received from our survey participants, three ways of engagement were identified: (1) building a community around the CC apps; (2) socialising with other CC users and (3) transparency with regard to the outcome/impact of a completed CC app task, i.e., being informed of the outcome of task after completion and the impact of it.

\begin{figure}[!t]
\centering
\includegraphics[width=0.28\textwidth]{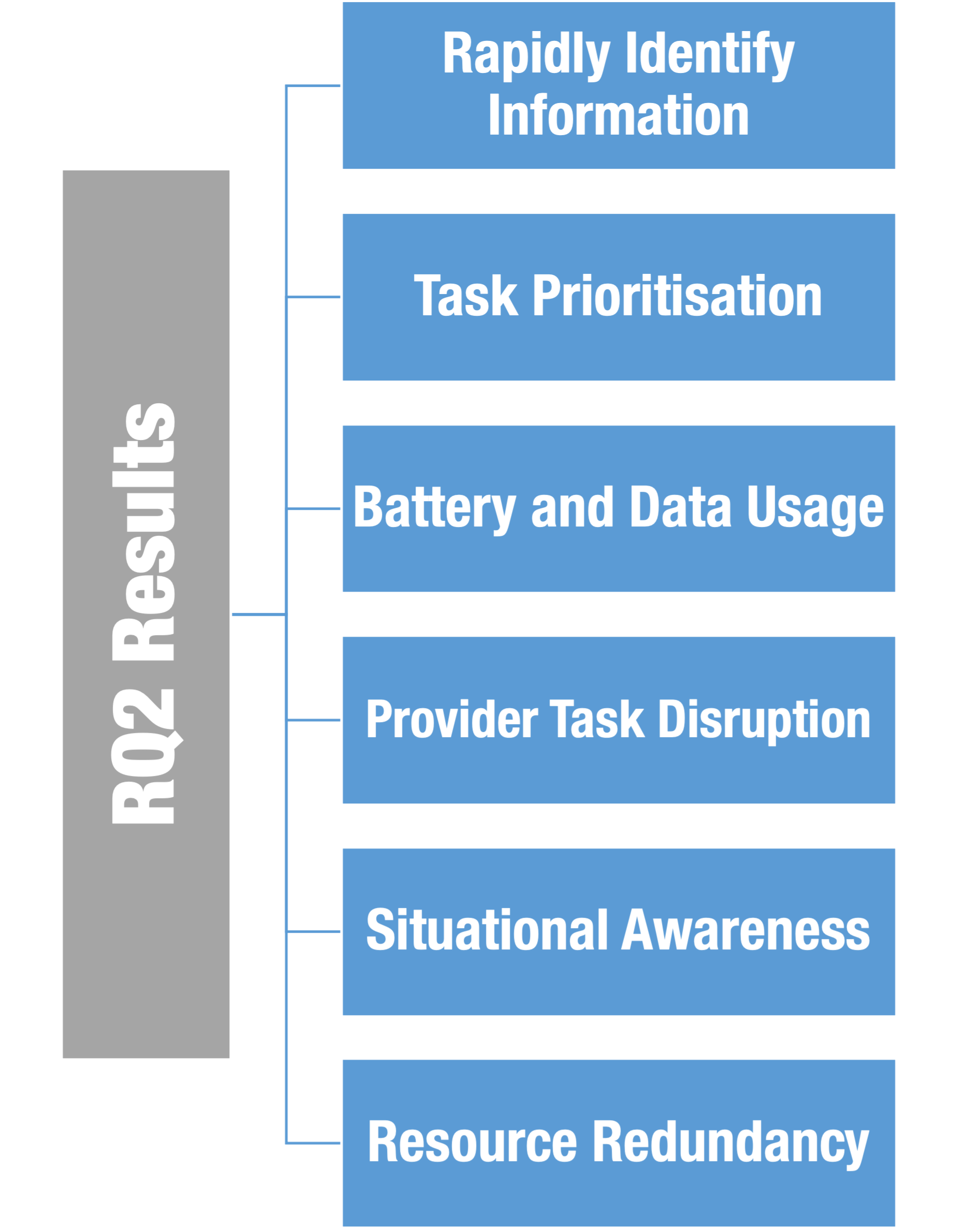}
\caption{Functional and Non-Functional Requirements from CC Apps}
\label{fig:rq2}
\vspace*{-1.5em}
\end{figure}

Fig.~\ref{fig:rq2} shows the results of our data analysis for RQ2. We note that RQ1 results also contribute to the functional and non-functional requirements of the overall CC infrastructure and CC app; however, RQ2 focuses directly on the requirements for CC apps.
The participants mentioned that one of their key considerations is quickly finding all the information regarding the tasks, intended outcomes, and the benefits of participation. On similar lines, they showed their inclination towards including the options of prioritising and sorting tasks based on their preferences and their history of working with tasks. The participants raised concerns about the battery and data usage of the app and the disruption of other applications running on their devices, previously known in the RE literature~\cite{bano2021rise,fazzini2022characterizing}. The participants further commented on the need for situational awareness in CC apps, as the change in the situation of consumers or providers could mean changes in the resources required, used or available. For instance, the resource capacity available for the same task in a café might be different from the resources in a moving train. The participants' last requirement in the survey was wasting resources or redundant usage. 

\section{Conclusion and Future Plans}
From what we observed in our survey, there are clearly human and social factors at play for CC to work at large, ranging from the usability of the software to incentive mechanisms, including monetary and motivations under the banner of ``doing good'' - analogous to ``AI for Good'', we have ``Crowd Computing for Good'' which can become a key motivation for many people providing their unused (or idle) computing resources. The ability to preserve privacy and reduce inconveniences (i.e., making participation in CC a seamless experience) for people will help to remove further hindrances. 

In future work, we will extend and build on this research to investigate how to embed human-centric SE aspects in the design of CC apps. 
First, the findings from this initial study on human factors and monetary and non-monetary incentives will be translated to formulate the set of features in a CC app, with further additions of human and social factors (such as age, gender, cultural background, disabilities, and emotions), incentive mechanisms including referrals and considering other theoretical aspects, such as the Elaboration Likelihood Model Theory (ELM)~\cite{petty2011elaboration}. We plan to develop high-fidelity CC mobile app prototypes with options to choose from different types of collaborative tasks appealing to different motivations. We plan to conduct a trial with $\approx$50 human participants, using the CC apps over four weeks, to evaluate the design. 
In this planned trial, we want to gather further data on RQ1 and RQ2 to corroborate our findings and experiment with different design features for addressing RQ3. For RQ3, we aim to study the design features that influence decisions by tapping into the human heuristics and biases that drive human reasoning~\cite{vallgaarda2012nudge}. The design decisions in apps can be in the form of pre-selected options, reminders, personalisation, framing and timing (e.g., sending notifications at a time when user is more likely to share their device) and use of social norms (e.g., disclosure of how one’s device sharing compares to one’s friends)~\cite{burr2018analysis}. For instance, these design decisions might work by modifying the content of a choice or the visualisation of a choice (e.g., in the design of the user interface)~\cite{schneider2018digital}.
The findings of this trial will contribute to developing a framework for CC apps development for different scenarios and application domains, which is the ultimate goal of this research project.

\bibliographystyle{IEEEtran}
\bibliography{nier22}

\end{document}